\documentclass[namedreferences]{solarphysics}
\usepackage[optionalrh]{spr-sola-addons} 
\usepackage{graphicx}        
\usepackage{color}           
\usepackage{url}             
\usepackage[pdfborder={0 0 0 },urlcolor=blue,breaklinks]{hyperref}
\ifx \arxivurl  \undefined \def \arxivurl#1{\href{http://arxiv.org/abs/#1}{\textsf{arXiv}}}\fi 
\ifx \doiurl    \undefined \def \doiurl#1{\href{http://dx.doi.org/#1}{\textsf{DOI}}}\fi 
\ifx \adsurl    \undefined \def \adsurl#1{\href{http://adsabs.harvard.edu/abs/#1}{\textsf{ADS}}}\fi 

\begin{document}

\sloppy

\begin{article}

\begin{opening}

\title{Latitudinal Distribution of Photospheric Magnetic Fields of Different
Magnitudes}

\author{E.S.~\surname{Vernova}$^{1}$\sep
        M.I.~\surname{Tyasto}$^{1}$\sep
        D.G.~\surname{Baranov}$^{2}$
       }
\runningauthor{E.S.~Vernova \textit{et al.}}
\runningtitle{Latitudinal Distribution of Photospheric Magnetic
Fields}

   \institute{$^{1}$ IZMIRAN, SPb. Filial, St.~Petersburg, Russian Federation
                     email: \url{helena@ev13934.spb.edu} \\
              $^{2}$ A.F.~Ioffe Physical-Technical Institute,
                          St.~Petersburg, Russian Federation
                     email: \url{d.baranov@mail.ioffe.ru} \\
             }

\begin{abstract}
Photospheric magnetic fields are studied using synoptic maps from
1976 to 2003 produced at the National Solar Observatory, Kitt Peak
(NSO/KP). Synoptic maps were averaged over the time interval of
nearly three solar cycles (Solar Cycles $21-23$). The change in
the latitudinal distribution was considered for the following
groups of magnetic field values: $B = 0 - 5$\,G, $B = 5 - 15$\,G,
$B = 15 - 50$\,G, and $B > 50$\,G.

Magnetic fields in each of the above groups have common
latitudinal distribution features, while for different field
groups these features change significantly.  Each of the groups is
closely related to a certain manifestation of solar activity.
Strong magnetic fields are connected with two types of solar
activity: active regions (magnetic fields $B>15$\,G) that are
related to sunspot zones,  and polar faculae (magnetic fields
$50\, {\rm G} > B > 15\, {\rm G}$) that occupy latitudes around
$65^\circ-75^\circ$. Fields from 5 to 15\,G occupy the polar
regions and are connected with polar coronal holes (the global
solar  dipole). Fields with $B<5$ G occupy a) the equatorial
region and b) latitudes $40^{\circ}-60^\circ$.

\end{abstract}


\end{opening}

\section{Introduction}
\label{intro} The distribution of solar activity over the surface
of the Sun and its change in the course of the 11-year solar cycle
represents one of the crucial points for the development of solar
dynamo models. The latitudinal distribution of  sunspots has a
long history of investigation; it is one of the most frequently
studied features of the solar cycle (see \opencite{hath}, and
references therein). Sp\"orer was one of the first researchers who
discovered the existence of the sunspot-generating zones and
described the regularity of their development. According to
Sp\"orer's law, discovered by R.C. Carrington (\opencite{carr}),
the mean latitude of sunspot groups gradually decreases from the
beginning to the end of the 11-year cycle of solar activity, {\it
i.e.}, the sunspot-generating zone moves from mid-heliolatitudes
toward the solar equator.

The characteristics of the latitude migration of sunspot groups in
the northern and southern hemispheres were investigated from 1874
to 1999 (\opencite{li}). It was found that the latitude
migration-velocity of sunspot groups is highest at the beginning
of a solar cycle, and as the solar cycle progresses, it decreases
with time, with an average of about $1.6^\circ \rm{yr}^{-1}$
during a solar cycle. Near minimum the centroid position of the
sunspot areas is about $28^\circ$ from the equator. The
equatorward drift ceases late in the cycle at about $7^\circ$ from
the Equator (\opencite{hath}). The sunspot zone latitudes and
equatorward drift measured relative to the starting time (near
solar minima) follow a standard path for all cycles that does not
depend upon the cycle strength or hemispheric dominance.

The width of the sunspot zone and its connection to the solar
cycle was studied in \inlinecite{mil1}, and \inlinecite{mil2}. The
active latitude bands are narrow at minimum, expand to a greatest
width at the time of maximum, and then narrow again during the
declining phase of the cycle. The latitude extension of the
sunspot-generating zone is closely related to the current level of
solar activity. The latitude distribution of solar activity from
1874 to 2004 was studied by \inlinecite{sola}, who calculated the
latitudinal moments of the sunspot group areas. Close
relationships between the total strength of the sunspot cycle, the
mean latitudes of the sunspots and the width of the sunspot zones
were found. A clear asymmetry was seen between the two
hemispheres: the range of variability from cycle to cycle in total
area, mean latitude, and width was less in the southern hemisphere
and the correlations between total area and mean latitude and
total area and width were stronger in the southern hemisphere.
According to \inlinecite{hath}, large-amplitude cycles reach their
maxima sooner than medium- or small-amplitude cycles (the
Waldmeier Effect; \opencite{wald1}, \citeyear{wald2}). Thus, the
sunspot zone latitude at the maximum of a large cycle will be
higher simply because the maximum occurs earlier and the sunspot
zones are still at higher latitudes.

Polar faculae present one more example of solar activity
concentrating around a certain range of heliolatitudes. Polar
faculae are visible on white-light images that have been obtained
daily at the Mount Wilson Observatory since 1906. Estimates of the
number of faculae in the vicinity of the north and south poles
were obtained for the interval 1906--1964 (\opencite{shee3}), and
then updated twice, first through 1975 and then through 1990
(\opencite{shee4}, \citeyear{shee5}). The third update with an
extensive review of measurement techniques and observed results
includes observations through the Spring of 2007 (see
\opencite{shee1}). The number of faculae at the north pole was
counted for  images obtained during the interval of 15 August --
15 September, when the north pole is most visible (for the south
pole the corresponding interval was 15 February -- 15 March).
Another procedure, which was adopted to determine the number of
the polar faculae, was used by \inlinecite{makarov} for images
selected from the Kislovodsk Solar Station collections
(1960--1994). The observed number of polar faculae was corrected
depending on the distance from the center of the disk and on the
season by means of empirical visibility functions. The visibility
functions for the north and south hemispheres were derived by
assuming that there is no seasonal variation in the distribution
of the polar-facula number for 1960--1994. The zone of the
emergence of polar faculae migrates poleward during the period
between the two successive polarity reversals of the solar
magnetic field  (\opencite{makarov2}). During the period from 1970
to 1978 the mean latitude of the zone where polar faculae emerge
increased from $55^\circ$ (1970) to $75^\circ$ (1978) in each
hemisphere. The last polar faculae were observed in the second
half of 1978 at latitudes from $70^\circ$ to $80^\circ$.

Polar faculae appear at higher latitudes than sunspots and precede
sunspots in their development by approximately six years
(\opencite{makarov}). Another study (\opencite{deng}) showed that
polar faculae lead the sunspot number by 52 months. Polar facular
measurements are in excellent agreement with polar magnetic flux
estimates and allow studying the evolution of the polar magnetic
field (\opencite{shee2}). A strong correlation between the
heliospheric magnetic field (HMF) and the polar flux at solar
minimum was found, whereas during periods of activity the HMF
showed a good correlation with the square root of the sunspot
area.

All manifestations of solar activity are closely connected to  the
magnetic field of the Sun and its cyclic changes. Butterfly
diagrams of the magnetic flux are a common way of studying the
large-scale evolution of the Sun's magnetism. In such a diagram,
the flux is averaged over a full rotation and plotted as a
function of the mid-time of the rotation and latitude.
\inlinecite{ulrich} averaged the magnetic field  as a function of
latitude over each solar rotation for 24 years of magnetic
observations from Mt. Wilson. The obtained diagram was used for
comparison with the dynamo model of \inlinecite{wang}. The
agreement between the model and the observations was quite
striking, although some deficiencies of the model were found.

The evolution of the zonal solar magnetic field from mid-1976 to
mid-1990 was studied by \inlinecite{hoeks}  using the Wilcox Solar
Observatory (WSO) magnetograms. The absolute value of the flux at
each latitude was averaged for each Carrington Rotation. The total
flux is closely related to the level of activity and resembles the
butterfly diagram of the sunspot location. At the solar minima the
total flux at all latitudes was low. About a year after minimum
the flux level rose at mid-latitudes and  expanded rapidly. The
peak-flux latitude gradually moved toward the equator in each
hemisphere. The polar flux was strongest near solar minimum
(\opencite{hoeks}). The features of variations over time,
latitude, and height of magnetic fields in $10^\circ$ latitudinal
zones of the Sun from the photosphere to the source surface were
studied by \inlinecite{akht}. A strong difference between
variations of the low- and high-latitude total fields was found.
At latitudes lower than $\pm 45^\circ$ in the large-scale field,
the contribution comes from magnetic fields of active regions,
plages and sunspots, in addition to the contribution of the
background fields. Variations over time of the total magnetic
field in the photosphere at low latitudes are similar to
variations of the Wolf numbers. In the polar regions the
contribution of strong fields is negligible. At latitudes of
$55^\circ$ and higher, the cyclical character of the variation of
the total field is small.

Magnetic fields of different magnitudes correspond to different
manifestations of the activity of the Sun. The following problems
are of interest: the study of the latitudinal distribution of the
solar magnetic fields for the whole range of magnetic field
strengths, the selection of magnetic field groups connected to
certain ranges of heliolatitude, and the comparison of these field
groups with different manifestations of the solar activity. A
novel feature of the present work is that we study the latitudinal
distribution of the photospheric magnetic fields obtained by
averaging the data over $1976-2003$. Thus, we do not consider the
time-dependency of the latitudinal distribution and study only
those characteristics that persist even after averaging over three
solar cycles.

In Section 2 we describe the data and discuss the method applied
in this article. Section 3 is devoted to the study of the
latitudinal distribution for groups of magnetic fields of
different strength. The connection of these groups to certain
manifestations of the solar activity is considered. In Section 4
the main conclusions are drawn.


\section{Data and Method}
\label{secdatameth}

For this study  we used synoptic maps of the photospheric magnetic
field produced at the NSO/KP (available at
\url{http://nsokp.nso.edu}). These data cover the period from 1975
to 2003 (Carrington Rotations $1625-2006$). Because the data have
many gaps during the initial  observation periods, we included the
data starting from Carrington Rotation 1646 in our analysis.
Synoptic maps have the following spatial resolution: $1^\circ$ in
longitude (360 steps), 180 equal steps in the sine of the latitude
from $-1$ (south pole) to $+1$ (north pole). Thus, every map
consists of $360\times 180$ pixels of magnetic-flux values.

On the assumption that the fields are radial to the solar surface,
the average line-of-sight component from the original observations
was divided by the cosine of the heliocentric angle (the angle
between the line of sight and the vertical direction) and
transformed into units of Gauss. Noisy values near the poles were
detected and replaced by a cubic spline fit to valid values in the
polar regions. Although the instrumental noise \textit{per}
resolution element is reduced by the map construction process, the
maps tend to be noisy at the poles because of noisy measurements
associated with the limb and the relatively small number of
original measurements that cover the poles. Some of the issues
related to the noise in the Kitt Peak measurements of the polar
magnetic field are discussed in  \inlinecite{murray}. Each year,
for the period of 1986--1990, the noise increased toward the limb.
The calculations of the noise level show that measurements in the
region with a viewing angle larger than $80^\circ$ cannot be
considered reliable. In the sine of the latitude representation of
the Kitt Peak data, only one latitude zone corresponds to the
range $80^\circ-90^\circ$. The NSO/KP data are complete in the
sense that where observational data are missing, such as when the
tilt of the Sun's axis to the ecliptic causes one of the polar
regions to be invisible, interpolated values are used
(\opencite{durr}).

Strong magnetic fields of both polarities occupy a relatively
small part of the Sun's surface. The magnetic-field strength for
the period $1976-2003$ shows a nearly symmetric distribution with
60.3\,\% of the strength values in the $0-5$\,G range, whereas
pixels with a magnetic strength above $50$\,G occupy only 3.3\,\%
of the solar surface. Magnetic fields in the $5-15$\,G and
$15-50$\,G strength ranges occupy 26.9\,\% and 9.5\,\%,
respectively.

\begin{figure}
\begin{center}
\includegraphics[width=0.75\textwidth]{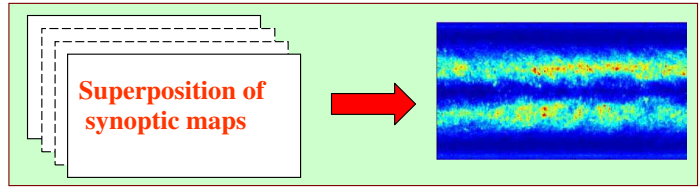}
\caption{Scheme of the method. Summing up synoptic   maps of the
photospheric magnetic field results in a synoptic map that
represents the distribution of the photospheric magnetic field
averaged over the entire period of $1976-2003$.} \label{summethod}
\end{center}
\end{figure}

We used the summation of synoptic maps over the period of nearly
three solar cycles (Cycles $21-23$) to obtain one averaged
synoptic map for the whole period under consideration. In
Figure~\ref{summethod} the scheme of the summation is presented.
By averaging the resulting map over the longitude, we obtain the
latitudinal profile of the magnetic flux for the period from 1976
to 2003. Our main goal is to compare the summary synoptic maps and
the latitudinal profiles for different groups of magnetic fields.

\begin{figure}
\begin{center}
\includegraphics[width=0.75\textwidth]{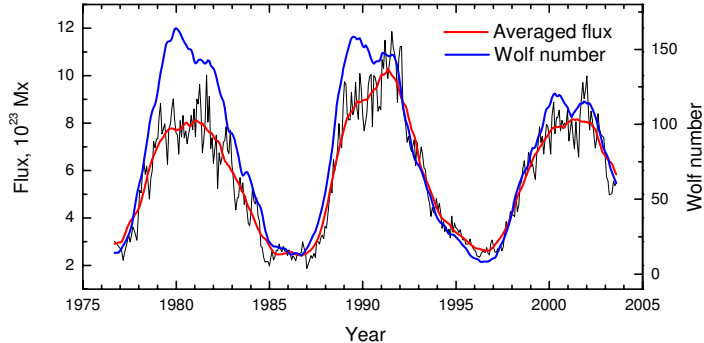}
\caption{Photospheric magnetic flux during Solar Cycles $21-23$
(thin black line). The red line represents smoothed moving
averages over 20 Carrington Rotations. Wolf numbers are plotted
with the blue line. The total magnetic flux  approximately follows
the evolution of the sunspot number.} \label{totalflux}
\end{center}
\end{figure}

To give an idea of the characteristics of the data we used and of
the order of magnitude of the magnetic fluxes, we show  the change
of the magnetic flux for years $1976-2003$ in
Figure~\ref{totalflux}, evaluated for each synoptic map and then
averaged using running means over 20 rotations. The photospheric
magnetic flux changes in phase with the solar activity. The
maximum of the flux coincides with the second Gnevyshev maximum;
the maximum fluxes occur in  1981.4, 1991.6, and 2002.2.

\begin{figure}
\begin{center}
\includegraphics[width=0.75\textwidth]{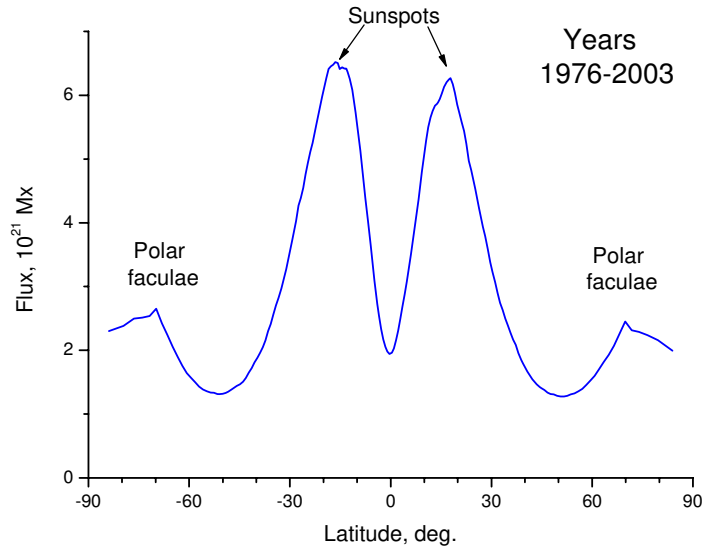}
\caption{Latitudinal profile of the magnetic field averaged over
Solar Cycles $21-23$. The magnetic flux was normalized to the
number of rotations included in the resulting map. Concentrations
of the magnetic field are observed in the sunspot zone and at the
latitudes characteristic of polar faculae.} \label{totalprofile}
\end{center}
\end{figure}

The averaged latitudinal profile of the magnetic field
(Figure~\ref{totalprofile}) shows two domains where the magnetic
flux increases: at the latitudes of the sunspot zone and at the
latitudes of the polar facular zone. The flux in the sunspot zone
 considerably exceeds the flux in the polar facular zone.

It should be mentioned that some very low (background) flux exists
at each latitude (approximately $1.3\times 10^{21}$\,G). The
maximum in the sunspot zone exceeds  the background flux by five
times. One can note that the total flux averaged over three cycles
for the southern hemisphere exceeds the flux in the northern
hemisphere.

\section{Results and Discussion}
\label{resdis}

When studying the latitudinal distributions of magnetic fields of
different magnitude, we discovered that there are four
characteristic groups of fields. These are fields with strengths
in ranges of $0-5$\,G, $5-15$\,G, $15-50$\,G, and $B>50$\,G.
Magnetic fields in each of the groups have common latitudinal
distribution features, while for different field groups these
features change significantly.  Each of the groups is closely
related to a certain manifestation of the solar activity.

To study the latitudinal distribution of these groups we obtained
separate summary magnetic field maps for each of the above field
ranges. Before the summation, each synoptic map was transformed in
such a way that only the pixels in the field strength range under
consideration are taken into account. Namely, the value of $B$ for
each pixel was replaced by 1 or 0 according to whether $B$ fell
into the $B_{min}-B_{max}$ range or not. Thus, we obtained the
summary map for the three solar cycles for the given range of
magnetic field strength. This map shows the percentage of time
when the fields of the given group at a given location were
present on the solar surface.

\begin{figure}
\begin{center}
\includegraphics[width=0.9\textwidth]{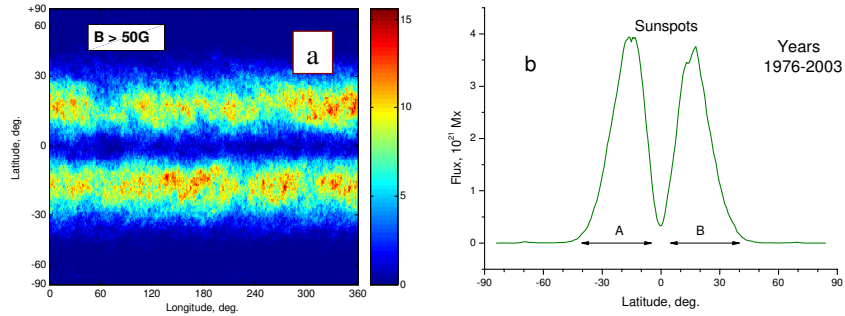}
\caption{Latitudinal distribution of strong magnetic fields
($B>50$\,G). a)~Summary magnetic field map for three solar cycles.
To the right we show a color bar indicating the percentage of time
for which the fields with strength $B>50$\,G were observed (the
maximum of the scale is $15\,\%$). b)~The latitudinal profile for
this group of fields shows that strong magnetic fields were only
present in the sunspot zones  (the latitudinal ranges are marked A
and B).} \label{strong}
\end{center}
\end{figure}

We show  the summary map for the group of the strongest magnetic
fields over 50\,G in Figure~\ref{strong}a.  These fields occupy
the sunspot zone and are obviously connected with the active
regions on the Sun. On the right,we give the scale that shows the
percentage of time when the fields from this group were observed
at each location. The maximum of the scale is about $15\,\%$. Dark
blue and blue correspond to rare appearance of the magnetic fields
in the given strength range (less than $5\,\%$ of time), while
yellow and red correspond to a frequent presence of the field
strength range considered ($10-15\,\%$ of the observation time).
By averaging the summary map over the longitude, we obtained the
latitudinal profile of the magnetic flux (Figure~\ref{strong}b).
The magnetic flux was normalized to the number of rotations
included into the resulting map. The latitudinal profile for this
group of field strengths shows that strong magnetic fields are
only observed in the sunspot zones.

\begin{figure}
\begin{center}
\includegraphics[width=0.9\textwidth]{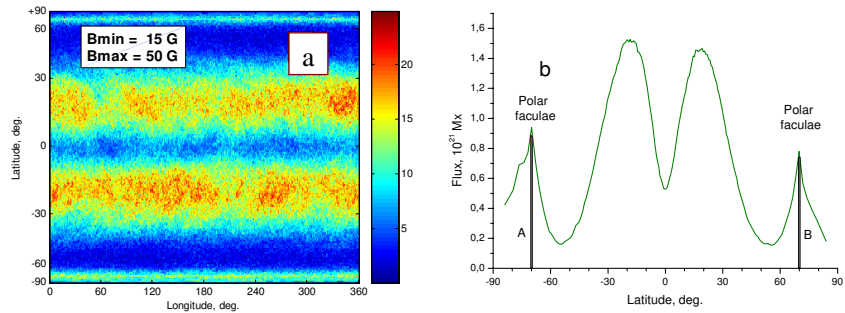}
\caption{Latitudinal distribution of magnetic fields in the
$15-50$\,G strength range. a)~Summary map for three solar cycles.
To the right we show a color bar indicating the percentage of time
for which the fields with strength $B=15-50$\,G were observed at a
given location (the maximum of the scale is $25\,\%$). b)~The
latitudinal profile of the magnetic flux. Dominating regions:
sunspot zone and zone of polar faculae (maxima of the polar
faculae magnetic flux are marked A and B). } \label{medium}
\end{center}
\end{figure}

In Figure~\ref{medium} the next group of magnitudes from 15 to
50\,G is considered. The summary map shows that each hemisphere
has two dominant regions corresponding to the sunspot zone and the
polar facular zone.  The sunspot zone occupies a wider strip of
latitudes than in Figure~\ref{strong}a (yellow and red strips at
latitudes below $40^\circ$) and exists  up to $25\,\%$ of the
time. The polar facular zone occupies a narrow strip (yellow and
light blue strips at latitudes around $70^\circ$), but it is
clearly visible. This zone is even more pronounced in the
latitudinal profile (Figure~\ref{medium}b), where the maxima of
facular concentration appear around latitudes $+70^\circ$ and
$-70^\circ$.

\begin{figure}
\begin{center}
\includegraphics[width=0.9\textwidth]{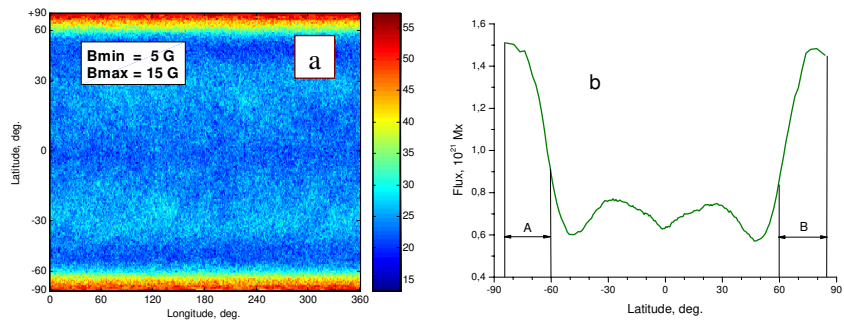}
\caption{Latitudinal distribution of magnetic fields in the
$5-15$\,G strength range. a)~Summary map for three solar cycles.
To the right we show a color bar indicating the percentage of time
for which the fields with strength $B=5-15$\,G were observed at a
given location (the maximum of the scale is $55\,\%$). b)~The
latitudinal profile of the magnetic flux. The strength and
localization of these fields indicate their connection with the
polar coronal holes (the latitudinal ranges are marked A and B). }
\label{weak}
\end{center}
\end{figure}

The group of fields with magnitudes from 5 to 15\,G dominates in
the summary map (Figure~\ref{weak}a) at the highest latitudes
(yellow and red strips in the map) and exists up to $55\%$ of
time. The lower limit of the color bar (dark blue color) in this
map corresponds to $15\,\%$ of time. Thus the magnetic fields of
this group appear over the whole solar disk for a significant part
of the solar cycle. The latitudinal profile of the magnetic flux
(Figure~\ref{weak}b) shows a sharp increase of the flux from
latitudes around $60^\circ$ toward the poles. The strength and
localization of these fields indicate their connection with the
polar coronal holes.

\begin{figure}
\begin{center}
\includegraphics[width=0.9\textwidth]{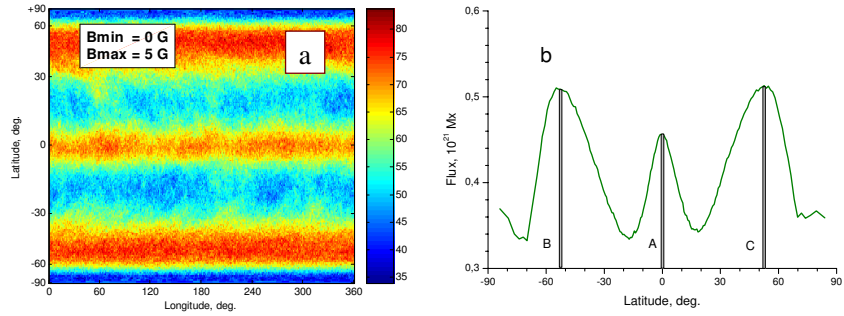}
\caption{Latitudinal distribution of magnetic fields in the
$0-5$\,G strength range. a)~Summary map for three solar cycles. To
the right we show a color bar indicating the percentage of time
for which the fields with strength $B=0-5$\,G were observed at a
given location (the maximum of the scale is $80\,\%$). b) The
latitudinal profile of the magnetic flux. The weakest fields are
concentrated at the solar equator (the flux maximum is marked A)
and within the latitude range $40^\circ - 60^\circ$ (the flux
maxima are marked B and C).} \label{veryweak}
\end{center}
\end{figure}

The weakest magnetic fields (lower than 5\,G) occupy three regions
in the summary map (Figure~\ref{veryweak}a): the equatorial region
$\pm 5^\circ$, and latitudes from $60^\circ$ down to the sunspot
zone in each of the hemispheres. As the color bar shows, weak
fields are present more than $80\%$ of time. At other latitudes
these fields appear at least for $35\,\%$ of time (dark blue
strips near the solar poles). The latitudinal profile of the flux
(Figure~\ref{veryweak}b) shows maxima at latitudes $0^\circ$ and
$53^\circ$.

\begin{figure}
\begin{center}
\includegraphics[width=0.9\textwidth]{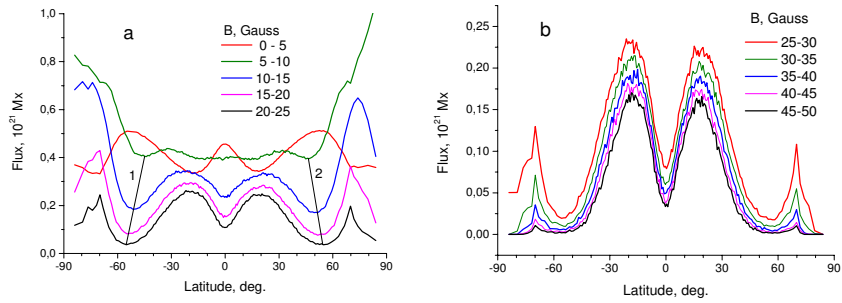}
\caption{Latitudinal profiles averaged over the period of
$1976-2003$ for field groups from 0 to 25\,G (panel a) and from 25
to 50\,G (panel b). Straight lines 1 and 2 in panel a show the
displacement of the profile minima.} \label{fivebin}
\end{center}
\end{figure}

The latitudinal distribution in Figure~\ref{strong} shows that
strong magnetic fields, $B > 50$~G, are confined only within the
sunspot zone. It is interesting to consider latitudinal
distributions of magnetic fields with strengths below 50 G for
5\,G strength bins. In Figure~\ref{fivebin}, we present the
latitudinal profiles of the magnetic field for various strength
ranges. The weakest (background) fields are concentrated near the
equator and at latitudes around $50^\circ$. In these regions
fields stronger than 10\,G are almost absent.

The high-latitude measurements are less reliable because the
magnetic field is radial in the photosphere and has only a small
component in the line-of-sight direction where the instrument is
sensitive. According to \inlinecite{harv},  the instrumental noise
level in the polar regions in NSO/KP maps is a function of
latitude and is on the order of  2 G  per map element. The
influence of these measurement errors is diminished by the data
treatment we used here. To obtain latitudinal profiles, synoptic
maps were averaged  in longitude and then averaged as a function
of time for nearly three cycles. With this averaging the
instrumental noise level near the poles is reduced significantly.

Latitudinal profiles for both $10-15$\,G and $15-20$\,G field
groups change in antiphase with the weakest magnetic fields, $B =
0 - 5$\,G. Flux minima for the range $15-20$\,G nearly coincide
with the weak-field maxima. The fluxes of magnetic field in
strength ranges of $B = 0 - 5$\,G and $B = 15 - 20$\,G, considered
as functions of the latitude, display a strong anticorrelation
(correlation coefficient $-0.94$).

In the polar regions the fields with strengths in the range
$5-10$\,G dominate; their concentration decreases sharply at
latitudes $\pm 60^\circ$ when we move from the poles toward the
equator. The magnetic flux in this range is nearly constant for
all heliolatitudes except the highest ones, where it exceeds the
flux values of all other field groups. For magnetic fields in the
range $10-15$\,G, the latitudinal profile has two maxima: in the
sunspot zone and in the zone of polar faculae. This feature
becomes more pronounced for higher magnetic strengths: $15-20$\,G
and $20-25$\,G. It should be noted that the flux in the zone of
polar faculae exceeds that of the sunspot zone for the weaker
magnetic fields of $10-15$\,G and $15-20$\,G, whereas for the
$20-25$\,G group we observe a reversed relation.

In Figure~\ref{fivebin}a the displacement of the minima toward
higher latitudes can be seen for the magnetic fields with higher
strengths. The minimum of the latitudinal distribution moves from
$45^\circ$ for the $5-10$\,G magnetic fields up to $55^\circ$
latitude for the $20-25$\,G fields. In contrast, the distribution
maxima in the sunspot zone move to the lower latitudes for greater
strengths, although this displacement is rather small.

As the strength of the magnetic field increases, two concentration
zones appear in the latitudinal distribution: the zone of polar
faculae (maximum at $\sim 70^\circ$) and the sunspot region
(maximum at $\sim 20^\circ$). This two-zone structure (sunspot
region and polar facular zone) is seen especially clearly in
Figure~\ref{fivebin}b for magnetic fields from $25-30$\,G to
$45-50$\,G. The flux in the sunspot zone significantly exceeds
that of the facular zone. Both fluxes decrease for greater
magnetic field strengths. The last group ($45-50$\,G) displays
high flux in the sunspot zone along with a flux close to zero  in
the facular zone.


\section{Conclusions}
\label{secconcl}

The latitudinal distribution of photospheric magnetic fields was
analyzed using synoptic maps of the NSO/KP ($1976-2003$). A close
connection between the values of the magnetic field and their
latitudinal localization was found, which persisted after
averaging the magnetic fields over three solar cycles.

The following groups of field values were shown to dominate at
different latitudes:

(a) from the equator to $5^\circ$, the weakest fields ($B=0 -
5$\,G);

(b) within the latitude range $5^\circ - 40^\circ$, the strong
fields ($B > 15$\,G), sunspots and active regions;

(c) in the latitude range $40^\circ  - 60^\circ$, the weakest
fields ($B=0 - 5$\,G);

(d) in a narrow strip of latitudes around $70^\circ$, magnetic
fields from 15 to 50\,G, polar faculae;

(e) in high-latitude regions (latitudes higher than $60^\circ $),
magnetic fields  from 5  to 15\,G, polar coronal holes.

The analysis of summary synoptic maps allowed us to distinguish
four characteristic groups of field values: $B = 0 - 5$\,G, $B= 5
- 15$\,G, $B = 15 - 50$\,G, and $B > 50$\,G. Within each of these
ranges, the magnetic fields have common latitudinal distribution
features, while for different field groups these features are
significantly different. For each of these field strength groups,
their latitudinal localization persists when we average  the
magnetic fields over three solar cycles. Each of the considered
groups of field values is closely related to a certain
manifestation of the solar activity.

\begin{acks}
NSO/Kitt Peak data used here are produced cooperatively by
NSF/NSO, NASA/GSFC, and NOAA/SEL. We thank the referee for many helpful comments.
\end{acks}

\end{article}


\begin{thebibliography}{}

\bibitem[\protect\citeauthoryear{Akhtemov {\it et al}.}{2015}]{akht}
Akhtemov, Z.S., Andreyeva, O.A., Rudenko, G.V., Stepanian, N.N.,
and Fainshtein, V.G.: 2015, {\it Adv. Space Res.} {\bf 55}, 3,
968. \adsurl{2015AdSpR..55..968A},
\doiurl{10.1016/j.asr.2014.09.036}.

\bibitem[\protect\citeauthoryear{Carrington}{1858}]{carr}
Carrington, R.C.: 1858, {\it Mon. Not. Roy. Astron. Soc.} {\bf
19}, 1, 1. \adsurl{1858MNRAS..19....1C}.

\bibitem[\protect\citeauthoryear{Deng {\it et al}.}{2013}]{deng}
Deng, L., Qu, Z., Dun, G., and Xu, C.: 2013, {\it Pub. Astron.
Soc. Japan} {\bf 65}, 1, 11. \adsurl{2013PASJ...65...11D},
\doiurl{10.1093/pasj/65.1.11}.

\bibitem[\protect\citeauthoryear{Durrant and
Wilson}{2003}]{durr} Durrant, C.J., Wilson, P.R.: 2003, {\it Solar
Phys.} {\bf 214}, 23. \adsurl{2003SoPh..214...23D},
\doiurl{10.1023/A:1024042918007}.

\bibitem[\protect\citeauthoryear{Harvey}{1996}]{harv} Harvey,
J.: 1996. \url {http://www.noao.edu/noao/staff/jharvey/pole.ps}.

\bibitem[\protect\citeauthoryear{Hathaway}{2015}]{hath}
Hathaway, D.H.: 2015, {\it Living Rev. Solar Phys.}, {\bf 12}, 4.
\url {http://www.livingreviews.org/lrsp-2015-4},
\doiurl{10.1007/lrsp-2015-4}. \adsurl{2015LRSP...12....4H}.

\bibitem[\protect\citeauthoryear{Hoeksema}{1991}]{hoeks}
Hoeksema, J.T.: 1991, {\it J. Geomagn. Geoelectr.} {\bf 43},
Suppl., 59.

\bibitem[\protect\citeauthoryear{Ivanov, Miletskii, and Nagovitsyn}{2011}]{mil1}
Ivanov, V.G., Miletskii, E.V., Nagovitsyn, Yu.A.: 2011, {\it
Astron. Reports} {\bf 55}, 10, 911. \adsurl{2011ARep...55..911I},
\doiurl{10.1134/S1063772911100040}.

\bibitem[\protect\citeauthoryear{Li, Yun, and Gu}{2001}]{li}
Li, K.J., Yun, H.S., Gu, X.M.: 2011, {\it Astron. J.} {\bf 122},
4, 2115. \adsurl{2001AJ....122.2115L}, \doiurl{10.1086/323089}.

\bibitem[\protect\citeauthoryear{Makarov and
Makarova}{1986}]{makarov2} Makarov, V.I., Makarova, V.V.: 1986,
{\it J. Astrophys. Astr.} {\bf 7}, 113.
\adsurl{1986JApA....7..113M}, \doiurl{10.1007/BF02715034}.

\bibitem[\protect\citeauthoryear{Makarov and Makarova}{1996}]{makarov}
Makarov, V.I., Makarova, V.V.: 1996, {\it
Solar Phys.} {\bf 163}, 267.
\adsurl{1996SoPh..163..267M},
\doiurl{10.1007/BF00148001}.

\bibitem[\protect\citeauthoryear{Miletsky, Ivanov, and Nagovitsyn}{2015}]{mil2}
Miletsky, E.V., Ivanov, V.G., Nagovitsyn, Yu.A.: 2015, {\it Adv.
Space Res.} {\bf 55}, 3, 780. \adsurl{2015AdSpR..55..780M},
\doiurl{10.1016/j.asr.2014.05.006}.

\bibitem[\protect\citeauthoryear{Mu\~noz-Jaramillo {\it et al}.}{2012}]{shee2}
Mu\~noz-Jaramillo, A., Sheeley, N.R., Zhang, J., DeLuca, E.E.:
2012, {\it Astrophys. J.} {\bf 753}, 2, 146.
\adsurl{2012ApJ...753..146M},
\doiurl{10.1088/0004-637X/753/2/146}.

\bibitem[\protect\citeauthoryear{Murray}{1992}]{murray} Murray,
N.: 1992, {\it Solar Phys.} {\bf 138}, 419.
\adsurl{1992SoPh..138..419M}, \doiurl{10.1007/BF00151926}.

\bibitem[\protect\citeauthoryear{Sheeley}{1964}]{shee3} Sheeley,
N.R., Jr.: 1964, {\it Astrophys. J.} {\bf 140}, 731.
\adsurl{1964ApJ...140..731S}, \doiurl{10.1086/147966}.

\bibitem[\protect\citeauthoryear{Sheeley}{1976}]{shee4}
Sheeley, N.R., Jr.: 1976, {\it J. Geophys. Res.} {\bf 81}, 3462.
\adsurl{1976JGR....81.3462S },
\doiurl{10.1029/JA081i019p03462}.

\bibitem[\protect\citeauthoryear{Sheeley}{1991}]{shee5}
Sheeley, N.R., Jr.: 1991, {\it Astrophys. J.} {\bf 374}, 386.
\adsurl{1991ApJ...374..386S},
\doiurl{10.1086/170129}.

\bibitem[\protect\citeauthoryear{Sheeley}{2008}]{shee1}
Sheeley, N.R., Jr.: 2008, {\it Astrophys. J.} {\bf 680}, 1553.
\adsurl{2008ApJ...680.1553S},
\doiurl{10.1086/588251}.

\bibitem[\protect\citeauthoryear{Solanki, Wenzler, and Schmitt}{2008}]{sola}
Solanki, S.K., Wenzler, T., Schmitt, D.: 2008, {\it Astron.
Astrophys.} {\bf 483}, 623. \adsurl{2008A&A...483..623S},
\doiurl{10.1051/0004-6361:20054282}.

\bibitem[\protect\citeauthoryear{Ulrich}{1993}]{ulrich} Ulrich,
R.K.: 1993, in  Weiss, W.W., Baglin, A. (eds.), {\it IAU Colloq.
137, Inside the Stars, Astron. Soc. Pac. CS} {\bf 40}, 25.
\adsurl{1993ASPC...40...25U}.

\bibitem[\protect\citeauthoryear{Waldmeier}{1935}]{wald1}
Waldmeier, M.: 1935, {\it Astron. Mitt. Z\"{u}rich} {\bf 14}, 105.
\adsurl{1935MiZur..14..105W}.

\bibitem[\protect\citeauthoryear{Waldmeier}{1939}]{wald2}
Waldmeier, M.: 1939, {\it Astron. Mitt. Z\"{u}rich} {\bf 14}, 470.
\adsurl{1939MiZur..14..470W}.

\bibitem[\protect\citeauthoryear{Wang, Sheeley, and
Nash}{1991}]{wang} Wang, Y.-M., Sheeley, N. R., Jr., Nash, A. G.:
1991, {\it Astrophys. J.} {\bf 383}, 431.
\adsurl{1991ApJ...383..431}, \doiurl{10.1086/170800}.

\end{thebibliography}
\end{document}